\documentclass[12pt]{article}

\usepackage{scicite}

\usepackage{graphicx}
\usepackage{amsmath}
\usepackage{times}
\usepackage{SIunits}
\usepackage{soul}
\usepackage{color}

\newcommand{\op}[1]{\hat{#1}}

\newcommand{\Eent}{\mathcal{E_{\text{Ent}}}}
\newcommand{\Eepr}{\mathcal{E}^{A\rightarrow B}_{\text{EPR}}}
\newcommand{\var}{\,\text{Var}}

\newenvironment{sciabstract}{%
\begin{quote} \bf}
{\end{quote}}

\title{Spatial entanglement patterns and Einstein-Podolsky-Rosen steering \\ in a Bose-Einstein condensate}

\author
{Matteo Fadel, Tilman Zibold, Boris D\'ecamps, Philipp Treutlein$^{\ast}$\\
\\
\normalsize{Department of Physics, University of Basel,}\\
\normalsize{Klingelbergstrasse 82, 4056 Basel, Switzerland}\\
\\
\normalsize{$^\ast$To whom correspondence should be addressed; E-mail: philipp.treutlein@unibas.ch}
}

\date{}

\begin{document} 


\baselineskip24pt


\maketitle



\begin{sciabstract} 
Many-particle entanglement is a fundamental concept of quantum physics that still presents conceptual challenges. While spin-squeezed and other non-classical states of atomic ensembles were used to enhance measurement precision in quantum metrology, the notion of entanglement in these systems was debated because the correlations between the indistinguishable atoms were witnessed by collective measurements only. 
Here we use high-resolution imaging to directly measure the spin correlations between spatially separated parts of a spin-squeezed Bose-Einstein condensate. We observe entanglement 
that is strong enough for Einstein-Podolsky-Rosen steering: we can predict measurement outcomes for non-commuting observables in one spatial region based on corresponding measurements in another region with an inferred uncertainty product below the Heisenberg relation. 
This could be exploited for entanglement-enhanced imaging of electromagnetic field distributions and quantum information tasks beyond metrology.
\end{sciabstract}


\newpage

Two quantum mechanical degrees of freedom are entangled (nonseparable) if the quantum state of one cannot be described independently of the other. When measurements are performed on both, entanglement results in correlations between the outcomes. While entanglement can exist between any quantum degrees of freedom, the conflict with classical physics is particularly striking when the correlations are observed between measurement outcomes obtained in spatially separated regions. 
Einstein, Podolsky and Rosen pointed out \cite{Einstein1935} that if the correlations are sufficiently strong, local measurements in one region $A$ can apparently change the quantum state in a spatially separated region $B$, a scenario Schr\"{o}dinger named ``steering'' \cite{Schroedinger35}. 
The possibility of steering between spatially separated systems implies that quantum theory is in conflict with a local realist description of the world \cite{WisemanPRL07,ReidRMP2009}. 
In fact, steering allows an observer in $A$ to use her local measurement outcomes to predict the outcomes of non-commuting measurements in $B$ with uncertainties below the Heisenberg uncertainty relation for $B$. 
EPR steering has been extensively explored with optical systems\cite{ReidRMP2009}. Entanglement was observed between spatially separated atomic ensembles \cite{JulsgaardNATURE2001,ChouNature2005,MatsukevichPRL2006,SimonNatPhys2007} and between individually addressable atoms in optical lattices \cite{Islam2015,Fukuhara2015}, but EPR steering has not yet been achieved for more than two atoms \cite{HagleyPRL97}.
Demonstrating the EPR paradox with ensembles of massive particles is desirable as it puts quantum physics to a stringent test in a new regime of increasingly macroscopic systems \cite{ReidRMP2009}. 
Moreover, it opens up new perspectives for applications of such systems in quantum metrology and one-sided device-independent quantum information tasks, which exploit EPR steering as a resource \cite{Cavalcanti2017}.

Experiments with ultracold atomic ensembles recently made rapid progress and a variety of nonclassical states can be prepared \cite{PezzeRMP2016}. Besides being of fundamental interest, such states find applications in quantum metrology \cite{GiovannettiNatPhoton2011}, where the correlations between the constituent atoms are exploited to reduce the noise in atom interferometric measurements \cite{GrossNATURE2010,LouchetChauvetNJP2010,LerouxPRL2010,OckeloenPRL2013}. Because of the large number of atoms involved, it is usually not possible to address and detect the atoms individually. In the case of Bose-Einstein condensates (BECs), it is even impossible in principle: the atoms are identical particles that occupy the same spatial mode. Still, quantum correlations between them can be characterized with the help of witness observables that involve only collective measurements on the entire ensemble \cite{AmicoRMP2008,GuehnePR2009}. This approach has been used to reveal the presence of entanglement\cite{GrossNATURE2010,RiedelNATURE2010}, EPR correlations\cite{PeiseNATCOM2015}, and even Bell correlations\cite{SchmiedSCIENCE16} in a cloud of atoms. However, these non-classical correlations have not yet been observed directly by performing measurements on spatially separated subsystems. Moreover, several authors have questioned whether the concept of entanglement in systems of indistinguishable particles is fully legitimate and useful for tasks other than metrology \footnote{For a brief review of the debate, see \cite{KilloranPRL2014}.}. 

As pointed out in the theoretical work of Killoran \textit{et al.} \cite{KilloranPRL2014}, the presence of entanglement in an ensemble of indistinguishable particles can be unambiguously confirmed by extracting it into spatially separated modes, turning it into a resource for a variety of quantum information tasks.
In our experiment, we demonstrate that entanglement can be extracted from spatially separated parts of a spin-squeezed BEC and use it to demonstrate the EPR paradox with an atomic system.

The quantum degrees of freedom in our experiment are two effective collective spins \cite{VuleticEffPRA15,PezzeRMP2016} $\op {\vec{S}}^A$ and $\op {\vec{S}}^B$ that describe the internal state of atoms in regions $A$ and $B$, respectively. Each atom is an effective two-level system with internal states $|1\rangle$ and $|2\rangle$. 
The component $\op S^{A}_{z} = \frac{1}{\eta_{\text{eff}}^{A}} \frac{(\op N^A_{1}-\op N^A_{2})}{2}$ is proportional to half the atom number difference between the states, evaluated in region $A$. The normalization by $\eta_{\text{eff}}^{A}$ \cite{VuleticEffPRA15}  takes into account the finite resolution of the imaging system (Supplementary Materials \cite{Supplementary}). A similar definition holds for $\op S^{B}_{z}$. Other spin components can be measured by applying appropriate spin rotations before detection.
To detect entanglement we use the criterion of Giovannetti \textit{et al.} \cite{GiovannettiPRA2003}, who have shown that for all separable states
\begin{equation}\label{eq:Eent}
\Eent = \dfrac{4\var(g_{z} \op S^{A}_{z} + \op S^{B}_{z}) \var(g_{y} \op S^{A}_{y} + \op S^{B}_{y})}{\left( \vert g_{z} g_{y} \vert \vert \langle \op S^{A}_{x} \rangle\vert + \vert \langle \op S^{B}_{x} \rangle\vert \right)^2} \geq 1 \; , 
\end{equation}
where $\var(\cdot)$ denotes the variance and $g_{z}$, $g_{y}$, are real parameters that can be optimized to minimize $\Eent$. Therefore, $\Eent < 1$ is a sufficient condition to certify entanglement (nonseparability) between $A$ and $B$. 

The variances in Eq.~\eqref{eq:Eent} quantify the uncertainty with which an observer in $A$ can predict (infer) the outcome of a spin measurement in $B$, based on a corresponding measurement on her own system, and are therefore called inferred variances. Since $\op S_{z}$ and $\op S_{y}$ do not commute, measuring both inferred variances requires repeated experiments on identically prepared systems. 

If correlations between $A$ and $B$ are strong enough, an observer in $A$ can predict the result of such non-commuting measurements performed by $B$ with a product of the inferred variances below the Heisenberg uncertainty bound for system $B$, \textit{i.e.} there is a violation of the relation \cite{ReidRMP2009}
\begin{equation}\label{eq:Eepr}
\Eepr = \dfrac{4\var(g_{z} \op S^{A}_{z} + \op S^{B}_{z})\var(g_{y} \op S^{A}_{y} + \op S^{B}_{y})}{ \vert \langle \op S^{B}_{x} \rangle\vert^2} \geq 1 \; .
\end{equation}
Note that if there are no correlations between $A$ and $B$, the variances in Eq.~\eqref{eq:Eepr} are minimized for $g_z=g_y=0$, for which the spin uncertainty relation for $B$ is recovered. 
In the presence of a violation of Eq.~\eqref{eq:Eepr}, $B$ must conclude that he is in the paradoxical situation considered by EPR, where $A$ is able to predict his measurement results without any classical communication.
Note that a violation of Eq.~\eqref{eq:Eent} does not imply a violation of Eq.~\eqref{eq:Eepr}, while the converse is true. This reflects the fact that entanglement is necessary but not sufficient for EPR steering, and that they are inequivalent types of correlations \cite{WisemanPRL07,QuintinoPRA15}. Moreover, the asymmetry between $A$ and $B$ present in Eq.~\eqref{eq:Eepr} implies that if $A$ can steer $B$ (denoted $A\rightarrow B$), then not necessarily $B$ can steer $A$ ($B\rightarrow A$), as investigated both theoretically \cite{Midgley2010,HePRL2015directional,QuintinoPRA15} and experimentally \cite{HaendchenNATPHYS2012,Wollmann2016} in optics.


To demonstrate a violation of both Eq.~\eqref{eq:Eent} and Eq.~\eqref{eq:Eepr} with a massive many-particle system, we perform experiments with two-component BECs of $N=590\pm 30$ ${}^{87}$Rb atoms, magnetically trapped on an atom chip \cite{BoehiNatPhys2009}. The two components correspond to the hyperfine states $\left \vert F = 1, m_{F}=-1 \right \rangle$ $\equiv \left \vert 1 \right \rangle$ and $\left \vert F = 2, m_{F}=1 \right \rangle \equiv \left \vert 2 \right \rangle$ and occupy nearly identical spatial modes. They can be described by a collective spin $\op {\vec{S}}$, referring to the entire BEC. 
We prepare the BEC in a spin-squeezed state by controlling atomic collisions with a state-dependent potential, as described in ref.~\cite{SchmiedSCIENCE16,RiedelNATURE2010,OckeloenPRL2013}. The spin-squeezed state features quantum correlations between the atoms, which reduce fluctuations of $\op S_z$ and increase fluctuations of $\op S_y$ while maintaining a large spin polarization in $\op S_x$ (see Fig.~\ref{fig:panel1}b). We obtain typically $\unit{-3.8(2)}{dB}$ of spin squeezing according to the Wineland criterion \cite{WinelandPRA1994}. 
Alternatively, we can prepare the BEC in a coherent spin state, where the atomic internal states are uncorrelated.

In order to access spatially separated regions in the BEC, we use the sequence illustrated in Fig.~\ref{fig:panel1}a.
After preparing the state, the atomic cloud is released from the trap and expands during a $\unit{2.2}{\milli \second}$ time-of-flight. This expansion is nearly spin-independent since collisional interactions are very similar for $\vert 1\rangle$ and $\vert 2\rangle$ and leads to a magnification of the atomic cloud.
Next, we set the axis $\vec{n}$ of the spin components $\op S_{\vec{n}}^A$ and $\op S_{\vec{n}}^B$ to be measured by applying a Rabi rotation pulse to the entire atomic cloud. 
Immediately thereafter, we record two high-resolution absorption images \cite{Ketterle99} of the atomic density distributions in states $|2\rangle$ and $|1\rangle$ by illuminating the atomic cloud twice with a resonant laser beam. The imaging pulses project the spin state and simultaneously localize the atoms in well-defined positions. Fig.~\ref{fig:panel1}c shows typical absorption images taken in this way. This experimental sequence is repeated thousands of times, alternating the measurement direction $\vec{n}$ along either $x$, $y$ or $z$.

We now define the two regions $A$ and $B$ to be analyzed on all pairs of absorption images (Fig.~\ref{fig:panel1}c). Counting the atom numbers $N^A_{1}$ and $N^A_{2}$ in region $A$ realizes a single-shot projective measurement of the local spin $\op S_{\vec{n}}^A = \frac{1}{\eta_{\text{eff}}^{A}} \frac{(\op N^A_{1}-\op N^A_{2})}{2}$. The same approach is applied to region $B$, which yields $\op S_{\vec{n}}^B$. 
The finite optical resolution and the motion of atoms during the imaging pulses amount to an uncertainty in the atomic position of $\unit{1.8}{\mu m}$ ($\unit{2.5}{\mu m}$) or $\unit{1.4}{pixels}$ ($\unit{2.0}{pixels}$) in the horizontal (vertical) direction. As a consequence, spins near the boundary have only partial overlap with the region $A$ or $B$, which is taken into account by $\eta_{\text{eff}}$
\cite{VuleticEffPRA15}. Furthermore, spins overlapping with both $A$ and $B$ lead to detection crosstalk, which we reduce by leaving a gap of $\unit{1}{pixel}$ between the two regions.  The experimentally determined spin variances include a contribution from detection noise, which can be independently measured and subtracted (see Supplementary Materials \cite{Supplementary} for details of the imaging method and data analysis).


To detect entanglement between regions $A$ and $B$ we evaluate Eq.~\eqref{eq:Eent} for different positions of the gap, corresponding to different splitting ratios $N^A/(N^A+N^B)$, where $N^A=N^A_1+N^A_2$ and similar for $N^B$ (Fig.~\ref{fig:entPanel}a, green dots). For a wide range of splitting ratios we observe a violation of the inequality in Eq.~\eqref{eq:Eent}, which goes far below the value that could be explained by detection crosstalk (Supplementary Materials \cite{Supplementary}). This proves that the two local spins $\op S^A$ and $\op S^B$ are entangled. The extracted entanglement derives from the quantum correlations among the indistinguishable atoms in the initial state\cite{KilloranPRL2014}, since the expansion of the cloud, the spin rotation and detection do not create the correlations.

An intriguing feature of our approach to extract entanglement \cite{KilloranPRL2014} from a many-body state is that the subsystems can be defined a posteriori on the images. This is in contrast to other experiments where the subsystems are defined by the experimental setup \cite{JulsgaardNATURE2001,ChouNature2005,MatsukevichPRL2006,SimonNatPhys2007} or by the source of the state \cite{ReidRMP2009,HaendchenNATPHYS2012}.
We exploit this feature to detect entanglement between regions $A$ and $B$ patterned in a variety of different shapes, see Fig.~\ref{fig:entPanel}b. The fact that we observe entanglement between all such regions reflects the symmetry of the underlying many-body quantum state: the quantum state of the indistinguishable bosons in the condensate has to be symmetric under particle exchange. Consequently, each atom is entangled with all other atoms, and the entanglement extends over the entire atomic cloud. 
In experiments with atoms in optical lattices, entanglement between different spatial bipartitions was observed by measuring entanglement entropy or concurrence, using systems of up to 10 atoms that were individually addressed \cite{Islam2015,Fukuhara2015}. By comparison, our experiment reveals entanglement in ensembles of hundreds of atoms using inequalities that apply in the continuous variable limit.


The correlations in our system are strong enough to demonstrate an EPR paradox: Fig.~\ref{fig:eprPanel}a shows a measurement of the EPR criterion Eq.~\eqref{eq:Eepr} for horizontal splitting of the cloud and different positions of the gap. We observe EPR steering $A\rightarrow B$ (green data points) for intermediate splitting ratios. For comparison, we evaluate the spin uncertainty relation $4\var(\op S^{B}_{z}) \var(\op S^{B}_{y})/ \vert \langle \op S^{B}_{x} \rangle\vert^2 \geq 1$ for system $B$, illustrating the reduction of the uncertainty product when replacing the non-inferred variances with the inferred ones. 
As can be seen in Eq.~\eqref{eq:Eepr}, EPR steering is an asymmetric concept. By relabeling region $A$ as $B$ and vice versa, we can invert the roles of the steering and steered systems. This inverted scenario also shows EPR steering $B\rightarrow A$ (red data points in Fig.~\ref{fig:eprPanel}a). The asymmetry between the curves indicates the presence of technical noise in our system \cite{HaendchenNATPHYS2012,WagnerJPB2014,HePRL2015directional}.
For intermediate splitting ratios we observe two-way steering $A\leftrightarrow B$, a prerequisite for observing the even stronger Bell correlations \cite{QuintinoPRA15}. We note that we also observe EPR steering if we do not subtract detection noise from the inferred variances, and also for vertical instead of horizontal splitting of the cloud.

Finally, we characterize the robustness of the observed EPR steering $A\rightarrow B$ to a variation of the gap size. We fix the central position of the gap such that the splitting ratio is $\unit{0.40}{}$ (the ratio maximizing steering $A\rightarrow B$ and $B\rightarrow A$ in Fig.~\ref{fig:eprPanel}a) and change the gap width symmetrically with respect to this position (Fig.~\ref{fig:eprPanel}c). We observe that EPR steering vanishes for large widths of the gap, where the size of the steered system is considerably reduced (Fig.~\ref{fig:eprPanel}b).


We have also performed measurements similar to Fig.~\ref{fig:entPanel} and Fig.~\ref{fig:eprPanel} with the BEC initially prepared in a coherent spin state, showing no statistically significant violations of Eqs.~\eqref{eq:Eent} and \eqref{eq:Eepr} beyond detection crosstalk (see Supplementary Materials \cite{Supplementary}). The observed spin noise in each individual region $A$ or $B$ agrees well with projection noise of uncorrelated atoms, confirming our calibration of the imaging system.


We have shown that entanglement and EPR steering can be observed between the collective spins in different spatial regions of a many-body system. These results are based on the extraction of entanglement from a system of identical particles, and on the observation that local collective spins, associated with arbitrary patterns in the atomic density images, satisfy criteria certifying entanglement and EPR steering. 
Our method can be used for quantum metrology of electromagnetic field patterns.
Consider an applied field that shifts the spin components $\op S^{B}_y$ and $\op S^{B}_z$ with respect to $\op S^{A}_y$ and $\op S^{A}_z$. The EPR entanglement allows one to detect this shift in the $yz$-plane with an uncertainty characterized by the product of the inferred variances in Eq.~\eqref{eq:Eepr}. The EPR parameter $\Eepr<1$ quantifies by how much this measurement improves over the Heisenberg uncertainty bound for $\op S^{B}$ and is thus a direct measure of the metrological enhancement provided by the EPR entanglement. Since our imaging method allows us to define the regions $A$ and $B$ a posteriori in a variety of shapes (see Fig.~\ref{fig:entPanel}), a single dataset could be used to analyze dipole, quadrupole and more complex patterns of the applied field. This is different from other field sensing methods where the pattern is defined by the state preparation \cite{GrossNATURE2010,Muessel2014}.

Beyond metrology, EPR steering is a resource for one-sided device-independent quantum information tasks \cite{Cavalcanti2017}. The asymmetry of the steering concept allows tasks such as quantum teleportation, entanglement swapping, or randomness certification to be performed in a situation where one of the involved parties can be trusted but not the other. An interesting perspective in this context is to distribute the correlations over macroscopic distances by splitting the atomic cloud with a double- or multi-well potential, exploiting the full control of BEC wavefunctions provided by the atom chip \cite{BoehiNatPhys2009}. 
Furthermore, our study raises the question whether Bell correlations could also be observed between spatially separated regions. While the EPR paradox can be demonstrated with Gaussian states and measurements and identical measurement settings in $A$ and $B$, a violation of a Bell inequality would require non-Gaussian states or measurements as well as the ability to measure different spin components in the two regions in a single run of the experiment \cite{BrunnerRMP2014}. This could be achieved by rotating the collective spins $\op S^{A}$ and $\op S^{B}$ independently with on-chip microwave near-fields, followed by atomic fluorescence detection with single-atom resolution. 
In summary, our results open up a variety of new perspectives for quantum science and technology with massive many-body systems.

Complementary to our work, the group of M. Oberthaler has observed spatially distributed multipartite entanglement and the group of C. Klempt has detected entanglement of spatially separated modes, \cite{OberthalerBB,KlemptBB}.

\bibliography{scibib}

\begin{thebibliography}{10}

\bibitem{Einstein1935}
A.~Einstein, B.~Podolsky, N.~Rosen, {\it Physical Review\/} {\bf 47}, 777
  (1935).

\bibitem{Schroedinger35}
E.~{Schr{\"o}dinger}, {\it Proceedings of the Cambridge Philosophical
  Society\/} {\bf 31}, 555 (1935).

\bibitem{WisemanPRL07}
H.~M. Wiseman, S.~J. Jones, A.~C. Doherty, {\it Phys. Rev. Lett.\/} {\bf 98},
  140402 (2007).

\bibitem{ReidRMP2009}
M.~D. Reid, {\it et~al.\/}, {\it Rev. Mod. Phys.\/} {\bf 81}, 1727 (2009).

\bibitem{JulsgaardNATURE2001}
B.~Julsgaard, A.~Kozhekin, E.~S. Polzik, {\it Nature\/} {\bf 413}, 400 (2001).

\bibitem{ChouNature2005}
C.~W. Chou, {\it et~al.\/}, {\it Nature\/} {\bf 438}, 828 (2005).

\bibitem{MatsukevichPRL2006}
D.~N. Matsukevich, {\it et~al.\/}, {\it Physical Review Letters\/} {\bf 96},
  030405 (2006).

\bibitem{SimonNatPhys2007}
J.~Simon, H.~Tanji, S.~Ghosh, V.~Vuletic, {\it Nature Physics\/} {\bf 3}, 765
  (2007).

\bibitem{Islam2015}
R.~Islam, {\it et~al.\/}, {\it Nature\/} {\bf 528}, 77 (2015).

\bibitem{Fukuhara2015}
T.~Fukuhara, {\it et~al.\/}, {\it Physical Review Letters\/} {\bf 115}, 035302
  (2015).

\bibitem{HagleyPRL97}
E.~{Hagley}, {\it et~al.\/}, {\it Physical Review Letters\/} {\bf 79}, 1
  (1997).

\bibitem{Cavalcanti2017}
D.~Cavalcanti, P.~Skrzypczyk, {\it Reports on Progress in Physics\/} {\bf 80},
  024001 (2017).

\bibitem{PezzeRMP2016}
L.~Pezz{\'e}, A.~Smerzi, M.~K. Oberthaler, R.~Schmied, P.~Treutlein,
  Non-classical states of atomic ensembles: fundamentals and applications in
  quantum metrology (2016).

\bibitem{GiovannettiNatPhoton2011}
V.~Giovannetti, S.~Lloyd, L.~Maccone, {\it Nature Photonics\/} {\bf 5}, 222
  (2011).

\bibitem{GrossNATURE2010}
C.~Gross, T.~Zibold, E.~Nicklas, J.~Est{\`e}ve, M.~K. Oberthaler, {\it
  Nature\/} {\bf 464}, 1165 (2010).

\bibitem{LouchetChauvetNJP2010}
A.~Louchet-Chauvet, {\it et~al.\/}, {\it New Journal of Physics\/} {\bf 12},
  065032 (2010).

\bibitem{LerouxPRL2010}
I.~D. Leroux, M.~H. Schleier-Smith, V.~Vuleti{\'c}, {\it Phys. Rev. Lett.\/}
  {\bf 104}, 250801 (2010).

\bibitem{OckeloenPRL2013}
C.~F. Ockeloen, R.~Schmied, M.~F. Riedel, P.~Treutlein, {\it Phys. Rev.
  Lett.\/} {\bf 111}, 143001 (2013).

\bibitem{AmicoRMP2008}
L.~Amico, R.~Fazio, A.~Osterloh, V.~Vedral, {\it Rev. Mod. Phys.\/} {\bf 80},
  517 (2008).

\bibitem{GuehnePR2009}
O.~G{\"u}hne, G.~T{\'o}th, {\it Physics Reports\/} {\bf 474}, 1 (2009).

\bibitem{RiedelNATURE2010}
M.~F. Riedel, {\it et~al.\/}, {\it Nature\/} {\bf 464}, 1170 (2010).

\bibitem{PeiseNATCOM2015}
J.~Peise, {\it et~al.\/}, {\it Nature Communications\/} {\bf 6}, 8984 EP
  (2015).

\bibitem{SchmiedSCIENCE16}
R.~Schmied, {\it et~al.\/}, {\it Science\/} {\bf 352}, 441 (2016).

\bibitem{KilloranPRL2014}
N.~Killoran, M.~Cramer, M.~B. Plenio, {\it Physical Review Letters\/} {\bf
  112}, 150501 (2014).

\bibitem{VuleticEffPRA15}
J.~Hu, W.~Chen, Z.~Vendeiro, H.~Zhang, V.~Vuleti\ifmmode~\acute{c}\else
  \'{c}\fi{}, {\it Phys. Rev. A\/} {\bf 92}, 063816 (2015).

\bibitem{Supplementary}
Materials and methods are available as supplementary materials on \emph{Science
  Online}.

\bibitem{GiovannettiPRA2003}
V.~Giovannetti, S.~Mancini, D.~Vitali, P.~Tombesi, {\it Phys. Rev. A\/} {\bf
  67}, 022320 (2003).

\bibitem{QuintinoPRA15}
M.~T. Quintino, {\it et~al.\/}, {\it Phys. Rev. A\/} {\bf 92}, 032107 (2015).

\bibitem{Midgley2010}
S.~L.~W. Midgley, A.~J. Ferris, M.~K. Olsen, {\it Physical Review A\/} {\bf
  81}, 022101 (2010).

\bibitem{HePRL2015directional}
Q.~Y. He, Q.~H. Gong, M.~D. Reid, {\it Phys. Rev. Lett.\/} {\bf 114}, 060402
  (2015).

\bibitem{HaendchenNATPHYS2012}
V.~H\"{a}ndchen, {\it et~al.\/}, {\it Nat Photon\/} {\bf 6}, 596 (2012).

\bibitem{Wollmann2016}
S.~Wollmann, N.~Walk, A.~J. Bennet, H.~M. Wiseman, G.~J. Pryde, {\it Physical
  Review Letters\/} {\bf 116}, 160403 (2016).

\bibitem{BoehiNatPhys2009}
P.~B{\"o}hi, {\it et~al.\/}, {\it Nat. Phys.\/} {\bf 5}, 592 (2009).

\bibitem{WinelandPRA1994}
D.~J. Wineland, J.~J. Bollinger, W.~M. Itano, D.~J. Heinzen, {\it Phys. Rev.
  A\/} {\bf 50}, 67 (1994).

\bibitem{Ketterle99}
W.~Ketterle, D.~S. Durfee, D.~M. {Stamper-Kurn}, {\it {{B}ose-{E}instein}
  condensation in atomic gases, Proceedings of the International School of
  Physics {``{E}nrico {F}ermi''}, Course {CXL}\/}, M.~Inguscio, S.~Stringari,
  C.~E. Wieman, eds. (IOS Press, Amsterdam, 1999), pp. 67--176.

\bibitem{WagnerJPB2014}
K.~Wagner, et. al., {\it J. Phys. B: At. Mol. Opt. Phys.\/} {\bf 47}, 225502
  (2014).

\bibitem{Muessel2014}
W.~Muessel, H.~Strobel, D.~Linnemann, D.~B. Hume, M.~K. Oberthaler, {\it
  Physical Review Letters\/} {\bf 113}, 103004 (2014).

\bibitem{BrunnerRMP2014}
N.~Brunner, D.~Cavalcanti, S.~Pironio, V.~Scarani, S.~Wehner, {\it Rev. Mod.
  Phys.\/} {\bf 86}, 419 (2014).

\bibitem{OberthalerBB}
Oberthaler paper.

\bibitem{KlemptBB}
Klempt paper.

\bibitem{reinaudi_strong_2007}
G.~Reinaudi, T.~Lahaye, Z.~Wang, D.~Gu\'{e}ry-Odelin, {\it Opt. Lett.\/} {\bf
  32}, 3143 (2007).

\bibitem{Joffe93}
M.~A. Joffe, W.~Ketterle, A.~Martin, D.~E. Pritchard, {\it J. Opt. Soc. Am.
  B\/} {\bf 10}, 2257 (1993).

\bibitem{McConnellNATURE2015}
R.~McConnell, H.~Zhang, J.~Hu, S.~{\'C}uk, V.~Vuleti{\'c}, {\it Nature\/} {\bf
  519}, 439 (2015).

\end{thebibliography}

\bibliographystyle{Science}

\section*{Acknowledgments}
\noindent
We thank M.~D. Reid, P.~D. Drummond, M.~Oberthaler, T.~Byrnes and R.~Schmied for useful discussions.
This work was financially supported by the Swiss National Science Foundation.
%
%
Author contributions: MF, TZ and BD performed experiments and analyzed data, supervised by PT. All authors discussed the results and contributed to the manuscript.

\section*{Supplementary materials}
Materials and Methods\\
Fig. S1, S2, S3, S4\\
References (41, 42, 43)

\clearpage
\newpage

\begin{figure}[h!]
	\begin{center}
	\includegraphics[width=\textwidth]{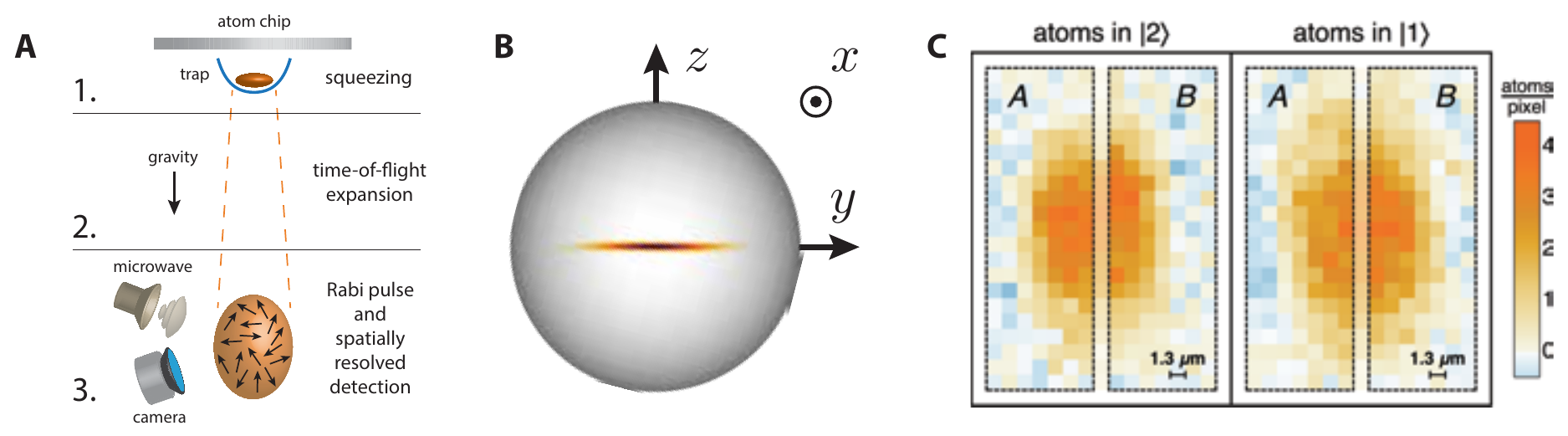}
	\caption{\label{fig:panel1}
		\textbf{Extracting entanglement from spatially separated regions of a BEC.}
		\textbf{a}: Experimental sequence. Step 1 consists in the preparation of a BEC in a spin squeezed state on an atom chip. In step 2 the trapping potential is switched off and the BEC expands. In step 3, a Rabi rotation pulse is applied to select the spin quadrature $\op S_{\vec{n}}$ to be measured, followed by recording two high-resolution absorption images of the atomic density distributions in states $\vert 1 \rangle$ and $\vert 2 \rangle$.
		\textbf{b}: Illustration of the spin-squeezed state on a sphere (Wigner function, representing the quantum fluctuations of the spin) and definition of the axes $\vec{n}$ used in the measurement of the entanglement and EPR steering criteria.
		\textbf{c}: Single-shot absorption images of the atomic densities in $\vert 2\rangle$ and $\vert 1\rangle$, showing example regions $A$ and $B$ used to define the collective spins $\op S^A$ and $\op S^B$ entering in the entanglement and EPR steering criteria. 
	}
	\end{center}
\end{figure}

\begin{figure}[h!]
	\begin{center}
	\includegraphics[width=\textwidth]{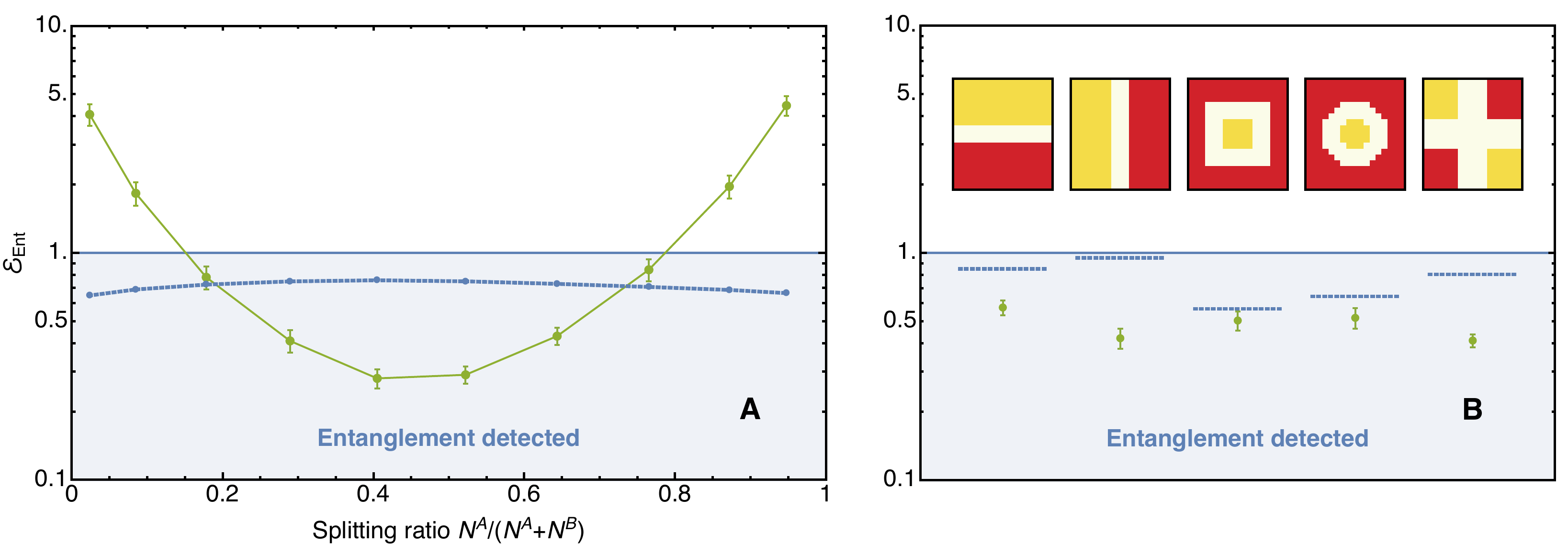}
	\caption{\label{fig:entPanel}
		\textbf{Spatial entanglement patterns in the atomic cloud. }
		\textbf{a:} Entanglement criterion Eq.~\eqref{eq:Eent} evaluated for a spin-squeezed BEC (green points) for different horizontal positions of the gap between regions $A$ and $B$ (see Fig.~\ref{fig:panel1}c), corresponding to different splitting ratios $N^A/(N^A+N^B)$. Lines are a guide to the eye and error bars indicate $\unit{1}{SEM}$. The blue points show the maximum violation that could be explained by detection crosstalk.
		\textbf{b:} Entanglement between regions of different shapes ($A$=yellow, $B$=red) in a spin-squeezed BEC. The pixel pattern used for the analysis is illustrated above the respective data points, and the blue segments show the corresponding maximum violation expected by crosstalk.
	}
	\end{center}
\end{figure}

\begin{figure}[h!]
	\begin{center}
	\includegraphics[width=\textwidth]{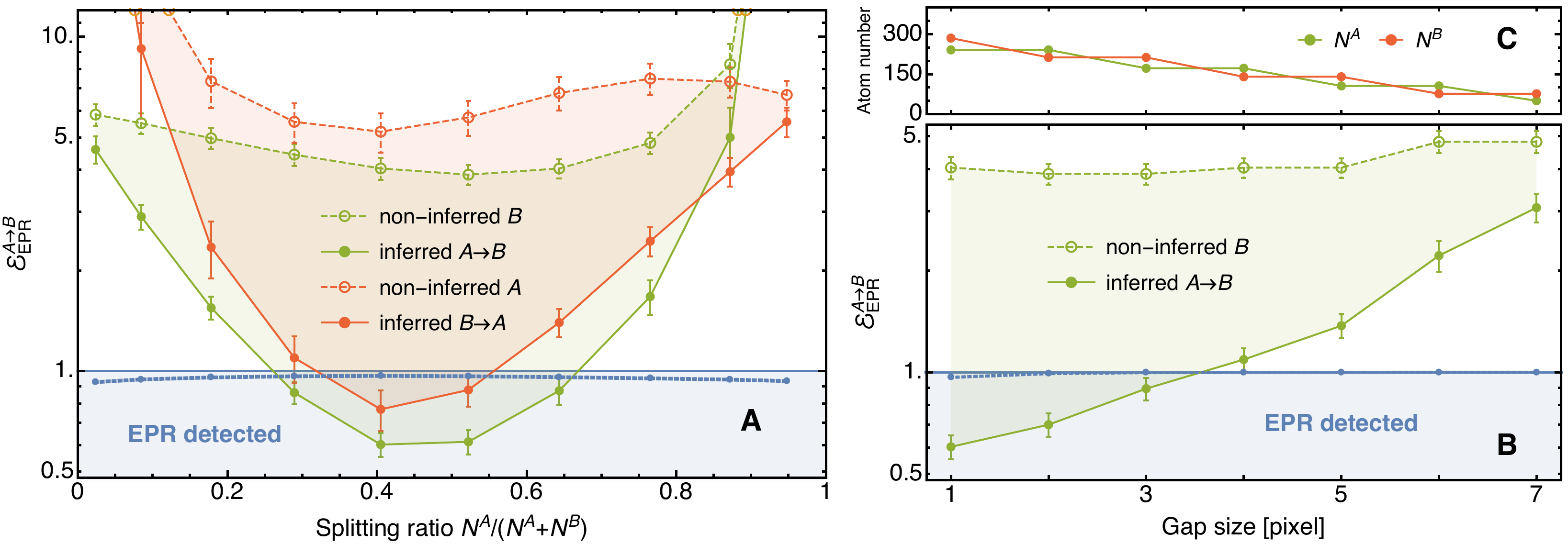}
	\caption{\label{fig:eprPanel}
		\textbf{Observation of Einstein-Podolsky-Rosen steering.}
		\textbf{a:} EPR steering criterion Eq.~\eqref{eq:Eepr}, evaluated for steering $A\rightarrow B$ (green filled circles) and $B\rightarrow A$ (red filled circles) in a spin-squeezed BEC, for different horizontal positions of the gap (see Fig.~\ref{fig:panel1}c), corresponding to different splitting ratios $N^A/(N^A+N^B)$. EPR steering is strongest for intermediate splitting ratios. Empty circles: spin uncertainty relation involving the product of non-inferred variances in region $B$ (green) and $A$ (red). Lines are a guide to the eye and the shaded regions are the reduction of the uncertainty product in replacing the non-inferred variances with the inferred ones. Blue points: maximum violation that could be explained by detection crosstalk.
		\textbf{b:} EPR steering $A\rightarrow B$ for different widths of the gap in Fig.~\ref{fig:panel1}c. The center of the gap is fixed to the position showing maximum EPR steering in Fig. \ref{fig:eprPanel}a for a width of one pixel. Even for increased gap size we find a significant violation of the bound, confirming that the correlations cannot be explained by detection crosstalk between the regions. Lines and shaded regions as in (\textbf{a}).
		\textbf{c:} Atom number in regions $A$ and $B$ as a function of the gap size.
	}
	\end{center}
\end{figure}


\clearpage
\newpage
\clearpage
\setcounter{figure}{0}
\renewcommand\thefigure{S\arabic{figure}}
\setcounter{equation}{0}
\renewcommand\theequation{S\arabic{equation}}
\setcounter{page}{1}

\section*{Supplementary Materials}

\subsection*{Experimental sequence description}
The preparation of the spin-squeezed state closely follows \cite{SchmiedSCIENCE16,OckeloenPRL2013,RiedelNATURE2010}. It starts with a BEC of $\approx 700$ atoms in state $ \left \vert 1 \right \rangle$. After a first $\pi / 2$ Rabi rotation, generating a coherent spin state, a state-selective potential produced by a microwave near field is turned on for $\unit{28}{\milli \second}$ at which time the resulting demixing-remixing dynamic reaches its first revival \cite{RiedelNATURE2010}. Then an echo $\pi$-pulse is applied before a second $\unit{28}{\milli \second}$ long state-selective potential period. The spin echo pulse is included to compensate for uncontrolled shot-to-shot phase shifts \textit{e.g.} due to the microwave potential and collisional interactions. In order to prevent uncontrolled offsets due to fluctuations of the Rabi rotation angle the spin-echo pulse phase is chosen such that the state is rotated around its mean spin value. With this sequence, we observe a spin squeezed state with typically $\unit{-3.8(2)}{dB}$ of spin squeezing according to the Wineland criterion.

In order to reduce the sensitivity of the spin-squeezed state to magnetic field fluctuations happening during its expansion, the state is rotated by $\unit{15}{}$ degrees to align it along the equator. Then, the magnetic trap is linearly ramped down in \unit{0.4}{ms}, and the cloud expands for \unit{2.2}{ms}. During time-of-flight, the magnetic field is kept, as well as possible, at its magic value $\approx \unit{3.23}{G}$ to reduce differential phase noise. Next, right before absorption imaging, Rabi rotations around orthogonal axes are performed in order to measure different spin components. For the $\pm \op S_{x}$ measurements, the state is rotated by $\pm \pi/2$ around the direction on the equator orthogonal to the mean spin. For the $\op S_{z}$ measurement, no rotation is applied, and for the $\op S_{y}$ measurement, a $\pi/2$ rotation around the mean spin direction is applied. The magnetic field is rotated in $\unit{0.2}{ms}$ to point along the imaging beam and reduced to $\sim \unit{1}{G}$. Finally, the absorption images are taken.

\subsection*{Imaging system and atom number calibration}
Our imaging system was previously described in detail in Ref.\cite{RiedelNATURE2010,OckeloenPRL2013} and we only recall here the parameters corresponding to the present experiments. We record two absorption images, taken \unit{1.33}{ms} apart, of the atomic population in the two internal states. Our detection system achieves atom number noise levels of $\sigma_{N_{1},\text{det}}=3.5$ atoms and $\sigma_{N_{2},\text{det}}=3.3$ atoms per whole picture. The effective scattering cross section $\sigma_{\text{eff}}$ is determined with the method described by Reinaudi \textit{et al.} \cite{reinaudi_strong_2007} and we find $\sigma_{\text{eff}}=(0.75\pm 0.02)\sigma_{0}$, where $\sigma_{0}$ is the scattering cross section of the cycling transition. Then, to take into account different detectivities of the two states, we perform Rabi oscillations with high contrast and ensure that the detected total atom number is independent of the relative population between the two states. Finally, to calibrate the absolute atom number, we observe the scaling of the projection noise with the total atom number. For a coherent state equally split in $\left \vert 1 \right \rangle$ and $\left \vert 2 \right \rangle$, we measure the variance of the relative atom number. We find that the projection noise dominates our measurement as the behavior is purely linear with a fitted slope of $1.04 \pm 0.005$. The small deviation from unity slope is then corrected for in our analysis.

\subsection*{Image analysis}
Due to the finite speed of the camera, the two absorption images are taken with a delay of \unit{1.33}{ms}. This means that the cloud of atoms in state $|1\rangle$ expands and falls longer than the cloud of atoms in state $|2\rangle$, leading to an increased rms size of $\sim 2\%$ (\unit{0.06}{pixel}) horizontally and $\sim 25\%$ (\unit{0.8}{pixel}) vertically (Fig.~\ref{fig:panel1}c), and to a center position which is $\sim \unit{140}{\micro m}$ lower. Note that already after the first image the spin state of the atoms is projected. However, after the first image only atoms in $\vert 2\rangle$ are spatially localized, while atoms in $\vert 1\rangle$ are still completely delocalized until the second image is taken. 

After all images have been recorded, we create two binary masks, one for state $\vert 1\rangle$ and the other for state $\vert 2\rangle$, defining the regions $A$ and $B$. Then, we evaluate the ensemble average (\textit{i.e.} average over all images) of the two atomic densities, and use it to center the two masks. Now that the two masks are defined and positioned, we apply the same masks to all individual pictures and count the atom numbers $N^A_1$, $N^A_2$, $N^B_1$ and $N^B_2$. Within the technical limitations (image resolution, blurring) discussed in the main text and in the following sections, our detection scheme realizes a projective measurement of the local collective spin in regions $A$ and $B$ of the expanded atomic cloud.

\subsection*{Optical resolution of the imaging system}
\begin{figure}[h]
	\centering
	\includegraphics[width=1\linewidth]{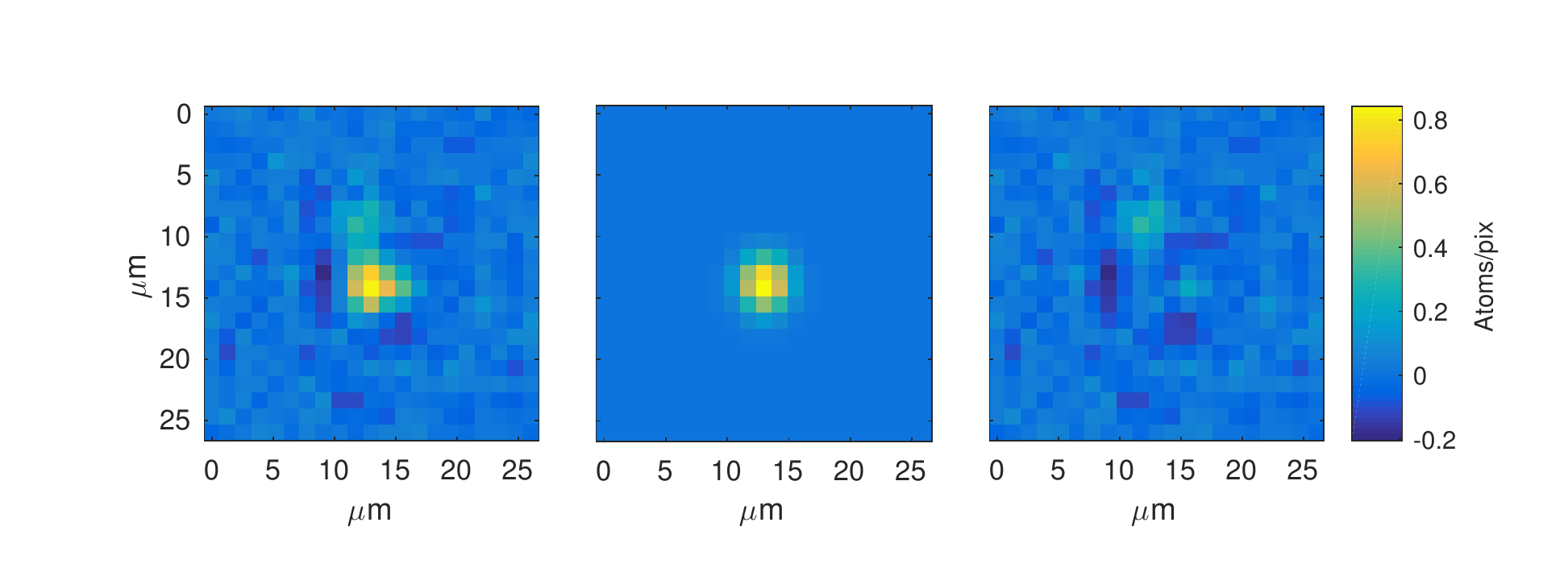}
	\caption{\label{fig:SmallCloud}
		The left panel shows an averaged absorption image of a small atomic cloud, taken a very short time after release from the trap. The gaussian fit (center panel) gives an upper bound of the size of the optical point spread function of our imaging system. In the right panel the fit residuals are shown. The colorbar applies to all three panels.}
	\label{figCSSDickeHusimi}
\end{figure}

To obtain an upper bound for the optical resolution of our imaging system, we image a small atomic cloud. 
To this end we prepare atoms in a trap which is approximately \unit{300}{\mu m} below the chip surface, such that the atoms are trapped at a position close to where the falling atomic cloud is in the actual experiments. We image the atomic cloud \unit{10}{\mu s} after switching off this trap, meaning that the atom density corresponds to a good approximation to the in-situ density. By using short laser pulses of \unit{10}{\mu s} for imaging, and averaging several absorption images of this small cloud, we obtain an upper estimate of the point spread function of our optical system. Figure \ref{fig:SmallCloud} shows the averaged absorption images, the Gaussian fit and the fit residuals. We find rms sizes of $\sigma_{\text{hor}}=\unit{1.1}{pixel}=\unit{1.43}{\mu m}$ and $\sigma_{\text{vert}}=\unit{1.2}{pixel}=\unit{1.56}{\mu m}$.

\subsection*{Image blurring due to random photon scattering during absorption imaging}
In absorption imaging the atomic cloud is illuminated by a pulse of resonant laser light. During the pulse the atoms scatter photons, which leads to a random velocity and position during the pulse. This blurring leads to a reduction of the effective optical resolution. In our experiment the pulse is very well described by a pulse of duration $\Delta t_\text{pulse}$ with constant intensity. We derive here a conservative estimate of the blurring. Since the atoms are mostly scattering photons on a cycling transition, we assume here a two-level model for the atomic transition. We further assume that the light is resonant during the whole imaging pulse. These assumptions overestimate the actual spread in position, since the real scattering cross section is smaller and the scattering rate is also reduced due to the longitudinal acceleration Doppler-shifting the atoms out of resonance during the pulse. These two effects are relatively small for our parameters, such that our estimate, although conservative, should still give reasonably good agreement with the experiment.  

We are interested in the transverse spread of position due to the random scattering. As derived by Joffe \textit{et al.} \cite{Joffe93}, the mean squared transverse position at time $t$ is given by 
\begin{align*}
x^2_\text{rms}(t)=\frac{1}{9} N_p(t) v^2_\text{rec}t^2=\frac{\Gamma}{18}\frac{s}{1+s}v^2_\text{rec}t^3 \;.
\end{align*}
Where $N_p(t)$ is the number of photons scattered between time $0$ and $t$, and $\Gamma$ and $s=\frac{I}{I_\text{sat}}$ are the decay rate and saturation parameter of the transition, respectively. This size is however only giving the rms transverse size of the atomic cloud at a given time. To estimate the rms size as observed on the image, we have to time-average the spatial distribution over the pulse length. To estimate this quantity, we consider a large number $M$ of atomic trajectories $x_j(t)$, $j= 1...M$. Then, the time averaged mean squared transverse position is
\begin{align*}
\bar{x}^2_\text{rms}(\Delta t_\text{pulse})&=\frac{1}{M}\sum_{j=1}^M\left[\frac{1}{\Delta t_\text{pulse}}\int^{\Delta t_\text{pulse}}_0x_j(t)\text{d}t\right]^2\leq\frac{1}{M}\sum_{j=1}^M\frac{1}{\Delta t_\text{pulse}}\int^{\Delta t_\text{pulse}}_0x^2_j(t)\text{d}t \;.
\end{align*}
In the last expression we can exchange the order of integral and sum and use the rms transverse size at time $t$ to estimate the expectation value of the set of trajectories. In this way we obtain
\begin{align*}
\bar{x}^2_\text{rms}(\Delta t_\text{pulse})&\leq\frac{1}{\Delta t_\text{pulse}}\int^{\Delta t_\text{pulse}}_0x^2_\text{rms}(t)\text{d}t=\frac{\Gamma}{72}\frac{s}{1+s}v^2_\text{rec}\left(\Delta t_\text{pulse}\right)^3 \;.
\end{align*}
If we take the estimate of the size obtained from in-situ absorption images (see above) into account and the blurring due to resonant absorption during the \unit{50}{\mu s} long imaging pulses, we obtain a total rms size of the blurred cloud on the camera of $\sigma_\text{blur}=\unit{1.4}{pixel}=\unit{1.8}{\mu m}$. We want to emphasize here that our estimation is conservative in the sense that it gives an upper bound for the blurring, overestimating the actual effect.

\subsection*{Resolution asymmetry due to atomic motion during probe pulse}

As the absorption images are taken during time of flight, the previously described imaging resolution and blurring, which describes the effective resolution in case of stationary atoms, needs to be averaged over the mean atomic trajectory during the imaging pulse. We do not detect any motion in the horizontal direction but atoms fall with a velocity in the vertical direction of $\unit{0.114(1)}{\mu m / \mu s}$ for the $F=2$ state and $\unit{0.129(1)}{\mu m / \mu s}$ for the $F=1$ state, the difference being compatible with gravitational acceleration between the two images. Integrated over the atom's trajectory during the imaging pulse, the total effective resolution keeps its Gaussian shape in the horizontal direction with a rms size of $\sigma_\text{hor}=\unit{1.4}{pixel}=\unit{1.8}{\mu m}$. In the vertical direction, it can still be well approximated by a Gaussian function as long as the falling distance during the imaging pulse is less than the full width at half maximum of the static blurred image. This results in state dependent effective rms sizes of the imaged cloud of $\sigma_\text{vert,F=2}=\unit{2.0}{pixel}=\unit{2.5}{\mu m}$ and $\sigma_\text{vert,F=1}=\unit{2.1}{pixel}=\unit{2.7}{\mu m}$ along the vertical direction.

\subsection*{Local collective spins}
\subsubsection*{1) naive definition}
In considering a region $A$ on the absorption images, we would be tempted to define the local collective spin as $\sum_{i \in A} \vec{s}_i^A$, where the index $i$ runs over all spins contained in $A$. However, in the case where the spins have some spatial extent, for example because of the point-spread function of the imaging system, the above definition of local collective spin needs to be corrected because the spin noise gets reduced and the usual spin commutation relations are not satisfied.
To see this, imagine $N$ atoms where the $i-$th has a density distribution $g(\vec{x},\vec{x}_i)$ centered at $\vec{x}_i$. A spin projection measurement corresponds to counting in a single experimental realization
\begin{equation}
\dfrac{N_1^A-N_2^A}{2} = \dfrac{1}{2} \int_A \left( \sum_{i=1}^{N_1} g(\vec{x},\vec{x}_i) - \sum_{i=N_1+1}^{N} g(\vec{x},\vec{x}_i) \right) \text{d}\vec{x} \;.
\end{equation}
Here, and in what follows, we assume for simplicity that the distribution $g(\vec{x},\vec{x}_i)$, the region $A$, and the distribution of the $\vec{x}_i$, are independent of the internal state of the atom. 
The spin expectation value is the ensemble average of this quantity over all possible positions $x_i$ and all possible partitions of $N$ into $N_1$ and $N_2=N-N_1$, denoted $\langle\cdot\rangle_{N_1}$.
If all the $\vec{x}_i$ come from the same probability density function $\rho$ we can write

{\footnotesize \begin{align}
\left\langle \left( \dfrac{N_1^A-N_2^A}{2} \right) \right\rangle &= \left\langle \int_{-\infty}^{+\infty} \rho(\vec{x}_1)\dots\rho(\vec{x}_N) \dfrac{1}{2} \int_A \left( \sum_{i=1}^{N_1} g(\vec{x},\vec{x}_i) - \sum_{i=N_1+1}^{N} g(\vec{x},\vec{x}_i) \right) \text{d}\vec{x} \text{d}\vec{x}_1 \dots \text{d}\vec{x}_N \right\rangle_{N_1} \nonumber\\
&= \left\langle \left( \dfrac{N_1-N_2}{2} \right) \int_{-\infty}^{+\infty} \rho(\vec{x}_1) \int_A g(\vec{x},\vec{x}_1)  \text{d}\vec{x} \text{d}\vec{x}_1 \right\rangle_{N_1} \nonumber\\
&= \left\langle \left( \dfrac{N_1-N_2}{2} \right) \right\rangle_{N_1} \int_{-\infty}^{+\infty} \rho(\vec{x}_1) f^A(\vec{x}_1) \text{d}\vec{x}_1  \label{eq:1moment}\;,
\end{align} }
where in going from the first to the second line we exchanged the order of the sum and the integrals, we used the fact that $\int_{-\infty}^{+\infty} \rho(\vec{x}_i)\text{d}\vec{x}_i = 1$, and that atoms can be relabeled. In the last line we introduced
\begin{equation}
f^A(\vec{x}_i) = \int_A g(\vec{x},\vec{x}_i)  \text{d}\vec{x} 
\end{equation}
representing the ``mode overlap'' between an atom centered in $\vec{x}_i$ and the region $A$. Similarly, for the total atom number in region $A$ we have
\begin{align}
\left\langle \left( N_1^A + N_2^A \right) \right\rangle &= \left\langle \int_{-\infty}^{+\infty} \rho(\vec{x}_1)\dots\rho(\vec{x}_N) \int_A \left( \sum_{i=1}^{N} g(\vec{x},\vec{x}_i) \right) \text{d}\vec{x} \text{d}\vec{x}_1 \dots \text{d}\vec{x}_N \right\rangle_{N_1} \nonumber\\
& = N  \int_{-\infty}^{+\infty} \rho(\vec{x}_1) f^A(\vec{x}_1) \text{d}\vec{x}_1  \label{eq:NtotalA}\;,
\end{align}
We now go through the same steps for the fluctuations. We consider the regions $U,V \in \{A,B\}$ and we calculate
{\footnotesize \begin{align}
& \langle \left(\dfrac{N_1^U-N_2^U}{2}\right)\left(\dfrac{N_1^V-N_2^V}{2}\right)  \rangle =  \\
&= \left\langle \dfrac{1}{4} \int_{-\infty}^{+\infty} \rho(\vec{x}_1)\dots\rho(\vec{x}_N) \times \right. \nonumber\\ 
&\phantom{space} \left. \times \int_U \left( \sum_{i=1}^{N_1} g(\vec{x},\vec{x}_i) - \sum_{i=N_1+1}^{N} g(\vec{x},\vec{x}_i) \right)  \int_V \left( \sum_{j=1}^{N_1} g(\vec{y},\vec{x}_j) - \sum_{j=N_1+1}^{N} g(\vec{y},\vec{x}_j) \right) \text{d}\vec{x}\text{d}\vec{y} \text{d}\vec{x}_1 \dots \text{d}\vec{x}_N \right\rangle_{N_1} \nonumber \\
& = \left\langle \dfrac{1}{4} \int_{-\infty}^{+\infty} \rho(\vec{x}_1)\dots\rho(\vec{x}_N) \times \right. \nonumber\\ 
&\phantom{space} \left. \times  \left( \sum_{i=1}^{N_1} f^U(\vec{x}_i) - \sum_{i=N_1+1}^{N} f^U(\vec{x}_i) \right)  \left( \sum_{j=1}^{N_1} f^V(\vec{x}_i) - \sum_{j=N_1+1}^{N} f^V(\vec{x}_i) \right) \text{d}\vec{x}_1 \dots \text{d}\vec{x}_N  \right\rangle_{N_1} \nonumber \\
& = \left\langle \dfrac{1}{4} \int_{-\infty}^{+\infty} \rho(\vec{x}_1)\dots\rho(\vec{x}_N) \left( \sum _{i=1}^{N}  f^{U}(\vec{x}_i) f^{V}(\vec{x}_i) +\sum_{i=N_1+1}^{N-1} \sum _{j=i+1}^{N} \left(f^{U}(\vec{x}_i) f^{V}(\vec{x}_j)+f^{U}(\vec{x}_j) f^{V}(\vec{x}_i)\right) + \right.\right. \nonumber\\
&\phantom{space} \left.\left. + \sum_{i=1}^{N_1-1} \sum _{j=i+1}^{N_1} \left(f^{U}(\vec{x}_i) f^{V}(\vec{x}_j)+f^{U}(\vec{x}_j) f^{V}(\vec{x}_i)\right) - \sum _{i=1}^{N_1} \sum _{j=N_1+1}^{N} \left(f^{U}(\vec{x}_i) f^{V}(\vec{x}_j)+f^{U}(\vec{x}_j) f^{V}(\vec{x}_i)\right)  \right) \text{d}\vec{x}_1 \dots \text{d}\vec{x}_N \right\rangle_{N_1} \nonumber \\
& = \left\langle \dfrac{1}{4} \int_{-\infty}^{+\infty} \rho(\vec{x}_1)\dots\rho(\vec{x}_N)  \Bigg( N f^{U}(\vec{x}_1)f^{V}(\vec{x}_1)+ \right. \nonumber\\ 
&\phantom{space} \left. + \dfrac{1}{2}\left( (N_1^2-N_1) + (N_2^2-N_2) - 2 N_1N_2 \right) \left(f^{U}(\vec{x}_1) f^{V}(\vec{x}_2)+f^{U}(\vec{x}_2) f^{V}(\vec{x}_1) \right) \Bigg) \text{d}\vec{x}_1 \dots \text{d}\vec{x}_N \right\rangle_{N_1} \nonumber \\
& = \dfrac{N}{4} \int_{-\infty}^{+\infty} \rho(\vec{x}_1) f^{U}(\vec{x}_1)f^{V}(\vec{x}_1) \text{d}\vec{x}_1 + \dfrac{1}{4} \left( \left\langle (N_1-N_2)^2 \right\rangle_{N_1} - N \right) \int_{-\infty}^{+\infty} \rho(\vec{x}_1)\rho(\vec{x}_2) f^{U}(\vec{x}_1) f^{V}(\vec{x}_2)  \text{d}\vec{x}_1 \text{d}\vec{x}_2  \label{eq:2moment} \;,
\end{align} }
which allows us to evaluate variances and covariances.

\subsubsection*{Example:}
In the case of a coherent spin state on the equator $\left\langle (N_1-N_2)^2 \right\rangle_{N_1} = N$ (this can be seen also from the fact that the average $\langle h(N_1) \rangle_{N_1}$ is given by the binomial distribution $\sum_{N_1=0}^N {{N}\choose{N_1}} p^{N_1} (1-p)^{N-N_1} h(N_1)$, with $p=1/2$.) This means that, from Eq.~\eqref{eq:NtotalA} and Eq.~\eqref{eq:2moment}, the ratio
\begin{equation}
\dfrac{\text{Var}(N_1^A-N_2^A)}{\langle N_1^A + N_2^A \rangle} = \dfrac{ \int_{-\infty}^{+\infty} \rho(\vec{x}_1) \left( f^A(\vec{x}_1)\right)^2 \text{d}\vec{x}_1}{ \int_{-\infty}^{+\infty} \rho(\vec{x}_1) f^A(\vec{x}_1) \text{d}\vec{x}_1} = \eta_{\text{eff}}^A \leq 1 \;,  \label{eq:VarOverNforA}
\end{equation}
where $\eta^A=1$ if and only if for all $\vec{x}_1$ where $\rho(\vec{x}_1)\neq 0$, $f^A(\vec{x}_1)$ is either 0 or 1. This can happen if either $A$ is the entire space, for which $f^A(\vec{x}_1)=1 \;\forall\; \vec{x}_1$, or $g(\vec{x},\vec{x}_1)=\delta(\vec{x}-\vec{x}_1)$, for which $f^A(\vec{x}_1)=1 \;\forall\; \vec{x}_1 \in A$. This means that a collective spins is properly defined if either we count atoms over all space or if atoms are properly localized.

We can evaluate Eq.~\eqref{eq:VarOverNforA} for our experimental situation taking $\rho$ to be the normalized BEC density for one of the states and $g(\vec{x},\vec{x}_1)$ the Gaussian point spread function of our imaging system. The analytical result is shown in Fig.~\ref{fig:figEta} together with experimental data. See also Ref. \cite{OberthalerBB}.

\subsubsection*{2) proper definition}
The problem illustrated by Eq.~\eqref{eq:VarOverNforA} has already been observed in experiments relying on atom-light interactions \cite{VuleticEffPRA15,McConnellNATURE2015}. In particular, it has been studied that atoms inside a cavity can couple with different interaction strengths to the optical mode, depending on their position. In this nonuniform scenario, the proper definition of the collective spin is given by $\vec{S} = \eta_{\text{eff}}^{-1} \sum_i \eta_i \vec{s}_i$, where $\vec{s}_i$ is the spin of the $i-$th particle, $\eta_i$ its coupling strength and $\eta_{\text{eff}}=\sum_i \eta_i^2/\sum_i \eta_i$. Following the same argument, we define our local spin as
\begin{equation}\label{EffSpin}
\vec{S}^A = \dfrac{1}{\eta_{\text{eff}}^A} \sum_{i} \eta_i^A \vec{s}_i \;,
\end{equation}
where $\eta_i^A = \int_A g(\vec{x},\vec{x}_i) \text{d}x = f^A(\vec{x}_i)$, and $\eta_{\text{eff}}^A = \langle (\eta^A)^2\rangle / \langle \eta^A \rangle$ is given by Eq.~\eqref{eq:VarOverNforA}. This definition allows us to define local spins for very small regions, in principle even of a single pixel size.

A spin projection measurement is given by
\begin{equation}\label{EffSpinProj}
S_{\vec{\alpha}}^A = \dfrac{1}{\eta_{\text{eff}}^A} \left(\dfrac{N_1^A-N_2^A}{2}\right) \;,
\end{equation}
where $\vec{\alpha}$ is the measurement axis along which the spin is projected and $N^A_{1,2}$ are the number of atoms counted in region $A$.

\begin{figure}[h!]
	\begin{center}
		\includegraphics[width=10 cm]{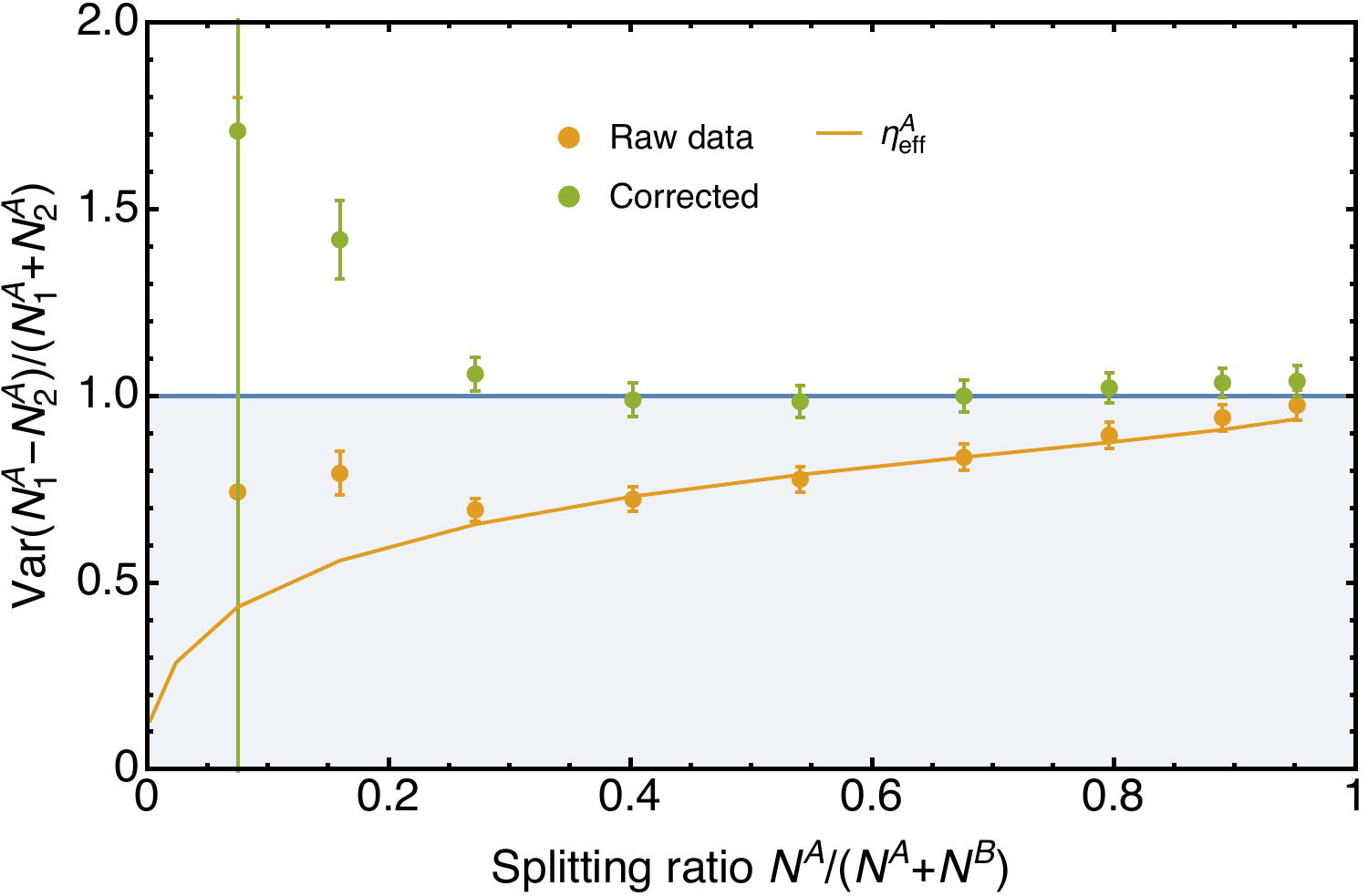}
	\caption{\label{fig:figEta}
		\textbf{Local spin fluctuations for a coherent spin state.}
		Normalized fluctuations of the local spin in region $A$ for different horizontal positions of the gap (see Fig.~\ref{fig:panel1}c), corresponding to different splitting ratios $N^A/(N^A+N^B)$. Orange dots: fluctuations evaluated using the measured atom number (raw data). For regions $A$ smaller than the entire cloud (splitting ratios $<1$) the fluctuations are suppressed by a factor $\eta_{\text{eff}}^A$ (orange line), as expected from Eq.~\eqref{eq:VarOverNforA}. Green dots: the correct definition of the local spin Eq.~\eqref{EffSpinProj} gives the expected fluctuations due to projection noise of uncorrelated atoms, Eq.~\eqref{corrVarOverN}. For region $B$ the behavior is similar. For very small regions (splitting ratios $<0.3$) technical noise increases the fluctuations above the coherent spin state projection noise.
	}
	\end{center}
\end{figure}

Note that with this definition, for a coherent state we have (using Eq.~\eqref{eq:VarOverNforA})
\begin{equation}\label{corrVarOverN}
\dfrac{4 \text{Var}\left(S_z^A \right)}{N} = \dfrac{\text{Var}\left((\eta_{\text{eff}}^A)^{-1}\left( N_1^A-N_2^A\right) \right)}{(\eta_{\text{eff}}^A)^{-1}\langle N_1^A + N_2^A \rangle} = \dfrac{\eta_{\text{eff}}^A}{\eta_{\text{eff}}^A} = 1 \;,
\end{equation}
independently on the definition of the region $A$. This is confirmed by comparison with the data in Fig.~\ref{fig:figEta}. We compute $\eta^{A}_{eff}$ from the mean detected atom densities by discrete numerical integration, according to Eq.~\eqref{eq:VarOverNforA}. Since the mean densities of the two spin states are slightly different, we use the one which gives the smallest $\eta^{A}_{eff}$ as a conservative value. The same approach is followed to compute $\eta^{B}_{eff}$.

\subsection*{Crosstalk}
\begin{figure}[h!]
	\begin{center}
		\includegraphics[width=10 cm]{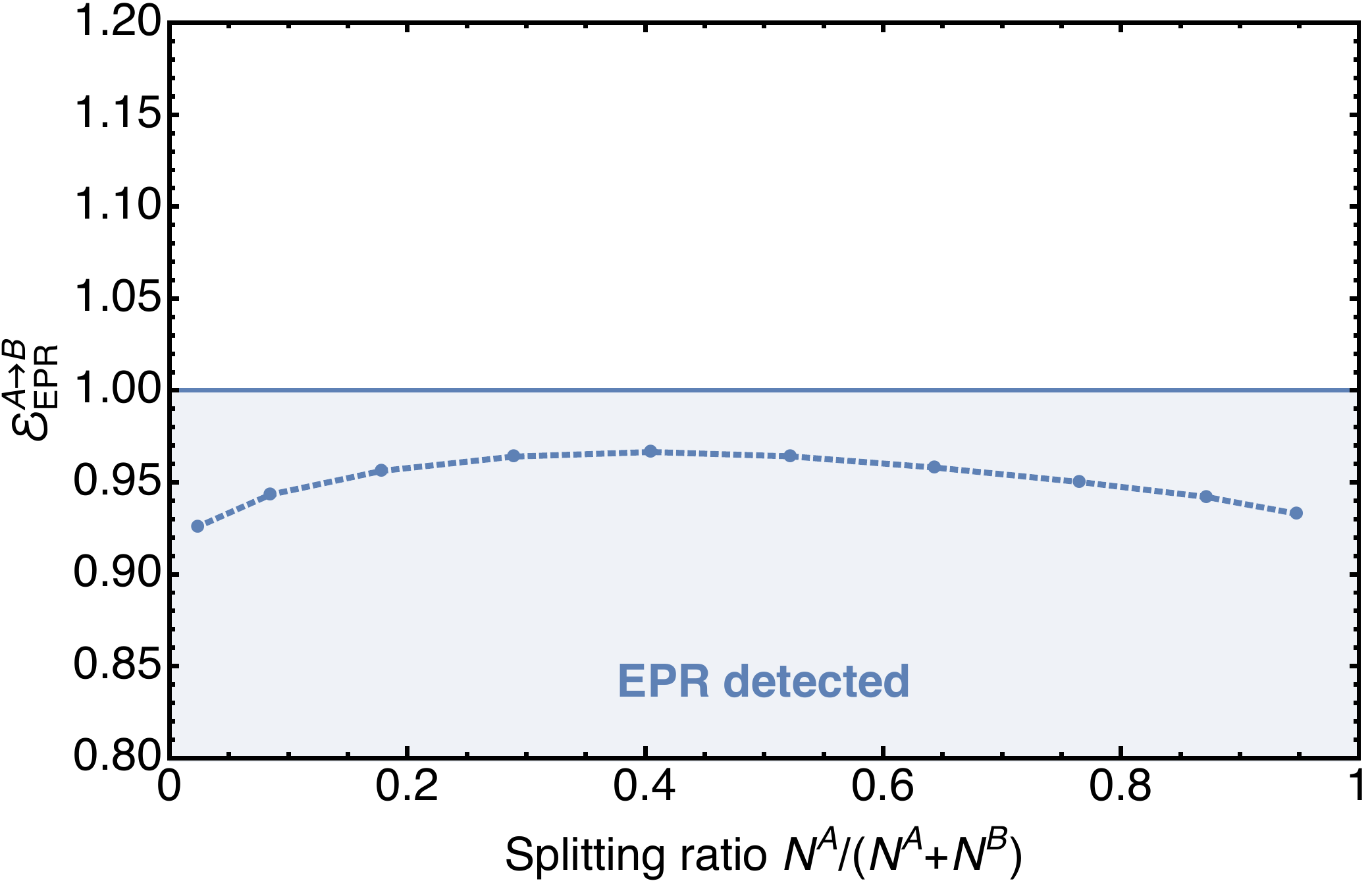}
	\caption{\label{fig:crosstalk}
		\textbf{EPR violation due to detection crosstalk between regions $A$ and $B$.}
		Crosstalk for different horizontal positions of the gap (see Fig.~\ref{fig:panel1}c), corresponding to different splitting ratios $N^A/(N^A+N^B)$. The blue dots correspond to Eq.~\eqref{anCrosstalk} divided by $\left( \eta_{\text{eff}}^B \right)^2$.
	}
	\end{center}
\end{figure}

Note that Eq.~\eqref{eq:2moment}, with $U=A$ and $V=B$, gives us also an estimation of the detection crosstalk between regions $A$ and $B$, which is due to atoms located at the boundary of $A$ and $B$ that contribute a signal to both regions. This crosstalk leads to classical correlations between the two regions, which could result in an apparent violation of the entanglement and EPR criteria. In particular, for the case of a coherent spin state with equal superposition we evaluate the EPR criterion for the optimal $g$'s
\begin{align} \label{anCrosstalk}
\dfrac{4 \left( \text{Var}(S_z^B)-\text{Cov}^2(S_z^A,S_z^B)/\text{Var}(S_z^A) \right) \left( \text{Var}(S_y^B)-\text{Cov}^2(S_y^A,S_y^B)/\text{Var}(S_y^A) \right)}{ \vert \langle S_x^B \rangle \vert^2 } = \hspace{3.5cm} \nonumber\\
= \dfrac{ \left( \left(\int_{-\infty}^{+\infty} \rho(\vec{x}_1) f^{A}(\vec{x}_1)f^{B}(\vec{x}_1) \text{d}\vec{x}_1\right)^2 - \int_{-\infty}^{+\infty} \rho(\vec{x}_1) \left( f^{A}(\vec{x}_1)\right)^2 \text{d}\vec{x}_1 \int_{-\infty}^{+\infty} \rho(\vec{x}_1) \left( f^{B}(\vec{x}_1)\right)^2 \text{d}\vec{x}_1 \right)^2 }{ \left(\int_{-\infty}^{+\infty} \rho(\vec{x}_1) \left( f^{A}(\vec{x}_1)\right)^2 \text{d}\vec{x}_1\right)^2 \left(\int_{-\infty}^{+\infty} \rho(\vec{x}_1) f^{B}(\vec{x}_1) \text{d} \vec{x}_1 \right)^2}  \;.
\end{align}
Note that this quantity is independent of $N$, and the asymmetry between $A$ and $B$ in the denominator. It is important to realize that an apparent violation of the EPR criterion can occur in the presence of crosstalk but also if the fluctuations in the regions are not properly estimated, which is already taken into account by the definition of collective spin Eq.~\eqref{EffSpin}. Since Eq.~\eqref{anCrosstalk} includes both effects, to isolate the apparent EPR violation caused by detection crosstalk, we divide it by $\left( \eta_{\text{eff}}^B \right)^2$.
For our experimental parameters we estimate that for a gap of one pixel, classical correlations originating from detection crosstalk between the two regions cannot decrease the EPR criterion below $0.94$ for the splitting ratios of our interest, see Fig.~\ref{fig:crosstalk}. This indicates that this effect is small compared to the observed EPR steering in Fig.~\ref{fig:eprPanel} of the main text.

The same idea has been applied to find the apparent violation of the entanglement criterion due to crosstalk. Since this criterion is much more sensitive to correlations between regions $A$ and $B$ than the EPR criterion, detection crosstalk gives a larger apparent violation in this case, see Fig.~\ref{fig:entPanel}. For the measurements performed on a spin-squeezed state, Fig.~\ref{fig:entPanel} and Fig.~\ref{fig:eprPanel}, the observed violation of the entanglement and EPR criteria cannot be explained by classical crosstalk between the regions.

\subsection*{Imaging noise subtraction}
In the experiment we measure $N_1^A + \delta_1^A$ and $N_2^A + \delta_2^A$, where $\delta_i^A$ is due to imaging noise in region $A$ for state $i$. Therefore, our measured effective spin is $\tilde{S}^A_\alpha = S^A_\alpha + \Delta^A$, where $\Delta^A = (\eta_{\text{eff}}^A)^{-1} ( \delta_1^A- \delta_2^A)/2$.
The variance entering the entanglement and the EPR criteria is
\begin{align}
\text{Var}\left(g_z \tilde{S}^A_z + \tilde{S}^B_z \right) &= \text{Var}\left(g_z (S^A_z + \Delta^A) + (S^B_z + \Delta^B) \right) = \nonumber\\
&= \text{Var}\left(g_z S^A_z + S^B_z \right) + \text{Var}\left(g_z \Delta^A + \Delta^B \right) + 2\text{Cov}(..., ...) \;,
\end{align}
where we assume the covariance to be zero because imaging noise is uncorrelated with the spin projection.
This allows us to subtract imaging noise from the criteria by writing the numerator as
\begin{align}\label{EPRnoImgNoise}
& \text{Var}\left(g_z S^A_z + S^B_z \right)\text{Var}\left(g_y S^A_y + S^B_y \right) = \nonumber\\
&\phantom{additionalspace} = \left( \text{Var}\left(g_z \tilde{S}^A_z + \tilde{S}^B_z \right) - \text{Var}\left(g_z \Delta^A + \Delta^B \right)  \right)\left( \text{Var}\left(g_y \tilde{S}^A_y + \tilde{S}^B_y \right) - \text{Var}\left(g_y  \Delta^A + \Delta^B \right)  \right) \;.
\end{align}
Note that the optimal $g$'s minimizing Eq.~\eqref{EPRnoImgNoise} are
\begin{equation}
g_z^\star = - \dfrac{\text{Cov}\left( \tilde{S}_z^A, \tilde{S}_z^B\right)}{\text{Var}(\tilde{S}_z^A)-\text{Var}(\Delta^A)} \;,
\end{equation}
and similarly for $g_y^\star$.

We estimate the contribution of imaging noise, $\text{Var}(\delta^A_i)$, from the photon shot noise on the actual absorption image. This estimate is conservative in the sense that additional noise sources such as dark counts, photon shot noise on the reference images and other sources of noise are neglected. However, in our case these sources give only a very small contribution to the noise. 

We obtain the conversion factor from the CCD counts to photon number by a calibration routine with flat-field correction. This number allows us to directly estimate the noise contribution of each pixel by the observed counts assuming Poissonian statistics of the photon numbers. By using error propagation this noise is converted into an effective atom number noise on the pixel.

For the subtraction of imaging noise from the observed spin variances we take the mean of this noise on the respective set of images.

\subsection*{Data acquisition and finite statistics}
The data acquisition is divided into small subsets to reduce the effect of possible drifts. In each subset, 4 measurements along $+x$ and 4 measurements along $-x$ give $\vert \langle \op S_{x} \rangle\vert$, followed by 70 measurements along $y$ and 60 measurements along $z$, which are used to compute variances. The regions A and B are defined for the whole data set while both entanglement (Eq.~\eqref{eq:Eent}) and EPR (Eq.~\eqref{eq:Eepr}) criteria are evaluated in each subset independently. Due to the finite statistics, care must be taken to use unbiased estimators. The unbiased sample estimate of inferred variances is:
\begin{equation}
\var\left(g_{z} \op S^{A}_{z} + \op S^{B}_{z}\right) = \frac{1}{m-2}\sum_{j=1}^{m}\left[\left(\bar{g_{z}} s^{A}_{z,j}+s^{B}_{z,j}\right)-\left(\bar{g_{z}} \bar{s^{A}_{z}}+\bar{s^{B}_{z}}\right)\right]^{2}
\end{equation}
where $m$ is the number of measurements, $\bar{g_{z}}$ is the optimal $g_{z}$ for the given subset, $s^{K}_{z,j} \left(K=A,B\right)$ are individual spin measurement results and  $\bar{s^{K}_{z}} \left(K=A,B\right)$ are the usual sample means. The normalization factor $m-2$ is necessary to take into account the additional degree of freedom originating from the inferring. This can be seen as the estimation of the variance of residuals in linear regressions.

To compute the entanglement and EPR criteria of the entire data set, we use the non-weighted mean of the subsets. Monte-Carlo simulation of our data analysis procedure allowed us to check that this results in unbiased means. 

Throughout all figures, the error bars we plot are the standard errors of the mean.

\subsection*{Coherent state measurements}
\begin{figure}[h!]
	\begin{center}
		\includegraphics[width=1\linewidth]{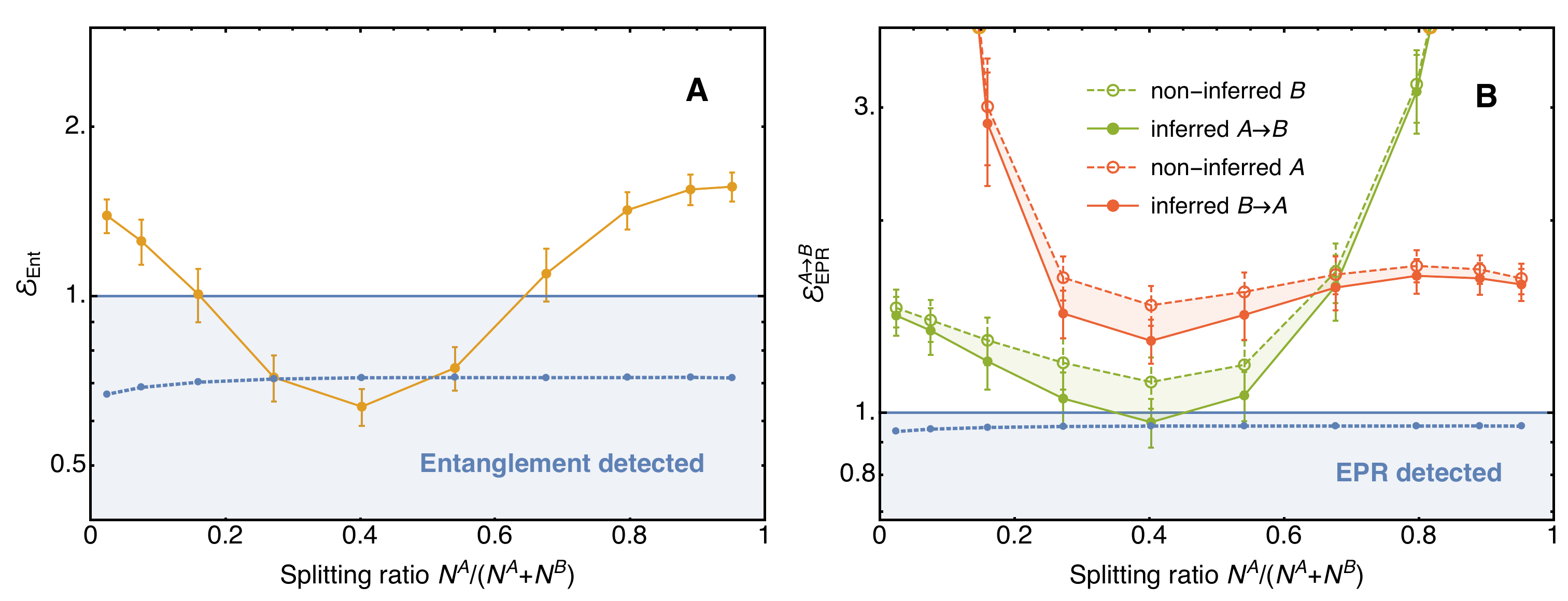}
	\caption{\label{fig:figEntEPRCSS}
		\textbf{Measurement of entanglement and EPR criteria for a BEC in a CSS.}
		\textbf{a:} Entanglement criterion Eq.~\eqref{eq:Eent} evaluated for a CSS and different horizontal positions of the gap between regions $A$ and $B$ (see Fig.~\ref{fig:panel1}c), corresponding to different splitting ratios $N^A/(N^A+N^B)$.
		\textbf{b:} EPR steering criterion Eq.~\eqref{eq:Eepr}, evaluated for steering $A\rightarrow B$ (green filled circles) and $B\rightarrow A$ (red filled circles) in a CSS, for different splitting ratios $N^A/(N^A+N^B)$. Empty circles: spin uncertainty relation involving the product of non-inferred variances in region $B$ (green) and $A$ (red). Lines are a guide to the eye and the shaded regions are the reduction of the uncertainty product in replacing the non-inferred variances with the inferred ones. Note that this reduction is much smaller than in the case of a spin-squeezed state, Fig.~\ref{fig:eprPanel}.
	}
	\end{center}
\end{figure}

In addition to the data for a spin squeezed state presented in the main text, we have performed measurements with a BEC in a coherent spin state (CSS). This state is prepared by a single $\pi/2$-pulse in the trap. In contrast to the spin squeezed state, the demixing-remixing sequence is omitted and the atoms are directly released from the trap. The readout of the different spin directions is done in the exact same way as in case of the spin squeezed state. 
The measured entanglement and EPR steering parameters we find for this state are shown in Fig.~\ref{fig:figEntEPRCSS}

The coherent state shows no EPR steering. It is also compatible with no entanglement for all splitting ratios, except one which does not have a very large statistical significance ($1.7$ sigma below the bound).

\end{document}